\documentclass[prl,twocolumn]{revtex4}
\usepackage{bm}
\usepackage{graphicx}
\usepackage{amssymb}
\usepackage{amsmath}
\usepackage{eufrak}
\usepackage{color}
\usepackage[utf8]{inputenc} 
\usepackage{hyperref}
\usepackage{pifont}
\usepackage{ulem}
\usepackage{epstopdf}

\newcommand{\nix}[1]{}

\newcommand{\MD}[1]{{\color{black} #1}}

\newcommand{\Mn}{\mathrm{Mn}}

\begin{document}


\title{Magnetooptical Study of Zeeman Effect in Mn modulation-doped InAs/InGaAs/InAlAs Quantum Well Structures}

\author{Ya.\,V.~Terent'ev$^{1,2}$, S.\,N.~Danilov$^1$, H.~Plank$^1$, J.~Loher$^1$,
D.~Schuh$^1$, D.~Bougeard$^1$, D.~Weiss$^1$, M.\,V.~Durnev$^2$,
S.\,A.~Tarasenko$^{2,3}$, I.\,V.~Rozhansky$^{2,3}$, S.\,V.~Ivanov$^2$, D.\,R.~Yakovlev$^{2,4}$, and
S.\,D.~Ganichev$^1$}
\affiliation{$^1$ Physics Department, University of Regensburg,
93040 Regensburg, Germany}
\affiliation{$^2$Ioffe Physical-Technical Institute,
194021 St.~Petersburg, Russia}
\affiliation{$^3$St.~Petersburg State Polytechnic University, 
195251 St. Petersburg, Russia}
\affiliation{$^4$ Experimentelle Physik 2, Technische Universit{\"a}t Dortmund,
44227 Dortmund, Germany}

\begin{abstract}

We report on a magneto-photoluminescence (PL) study of Mn modulation-doped
InAs/InGaAs/InAlAs quantum wells.
Two PL lines corresponding to the radiative recombination of photoelectrons with free
and bound-on-Mn holes have been observed.
In the presence of a magnetic field applied in the Faraday geometry
both lines split into two circularly polarized components.
While temperature and magnetic field dependences of the splitting
are well described by the Brillouin function, providing an evidence for
exchange interaction with spin polarized manganese ions, the value
of the splitting exceeds the expected value of the giant Zeeman splitting by two orders of magnitude
for a given Mn density. Possible reasons of this striking observation are discussed.
%
%
\end{abstract}
\pacs{xxx}

\maketitle{}

\section{I. Introduction}

Implementation of spintronics concepts requires
semiconductor heterostructures  with evident magnetic
properties.   To enhance the interaction between carrier spins and
a magnetic field and to achieve ferromagnetic spin ordering heavy
doping of materials with magnetic ions is required. Diluted
magnetic semiconductors (DMS) based on narrow-gap III-V compounds,
and in particular InMnAs, are considered to be promising
candidates for application due to a relatively high Curie
temperature and the strong spin-orbit interaction~\cite{Dietl2010,Schallenberg2006}.
While InAs-based DMS systems with strong spin-orbit coupling have
been realized and show very interesting
magnetotransport~\cite{Ohno92,Wurstbauer2010,Rupprecht} 
and opto-electronic
properties~\cite{Zudov2002,5aa,Terentyev2011,Olbrich2012,Khodaparast2013},
direct measurements of the giant Zeeman splitting by means of 
polarized magneto-photoluminescence (PL) have not been reported so far.
This is primary caused by the fact that
the doping of III-V compounds by Mn atoms generates numerous lattice defects (even if they are partially removed by annealing~\cite{Schallenberg2006})and, consequently, to a drastic decrease of the radiation efficiency.
This obstacle can be overcome in InAs quantum wells (QW)
in which the barrier is modulation-doped with Mn and Mn atoms are transported by segregation
into the InAs QW.

Here, we study the magneto-PL of high-quality DMS heterostructures
InAs/In$_{0.75}$Ga$_{0.25}$As/In$_{0.75}$Al$_{0.25}$As:Mn, where small concentration of Mn in the
InAs QW was provided by segregation. In zero magnetic field, the low-temperature PL is contributed by  
two lines separated by 25~meV. The analysis shows that the observed PL lines stem
from the recombination of free and bound-on-Mn  holes with
photoexcited electrons. A magnetic field applied in the Faraday geometry results in the splitting of each PL line
into two circularly polarized components with the opposite helicity.
The splitting is found to be a nonlinear function of the magnetic field and strong temperature dependent.
These findings are qualitatively well described by the presented theory which takes into account
radiative recombination processes in the InAs QW and the
exchange interaction of carriers with spin polarized Mn ions.
At the same time, the value of the splitting
exceeds the expected giant Zeeman splitting by two
orders of magnitude.
Possible reasons of this striking observation are discussed.

\section{II. Samples}

The InAs/In$_{0.75}$Ga$_{0.25}$As/In$_{0.75}$Al$_{0.25}$As QW
heterostructures investigated were
fabricated by molecular beam epitaxy (MBE) on a fully relaxed metamorphic
In$_{x}$Al$_{1-x}$As$/(001)$GaAs compositionally graded buffer where the
In content $x$ is increased stepwise from 0.05 to 0.75 over  1~$\mu$m of layer thickness.
Such an approach on structure design enables the fabrication of high quality
defect-free and strain relaxed virtual substrates of high indium content that allow an
effective collection of photogenerated carriers into the QW ~\cite{APL2014}.

\begin{figure}[b]
\includegraphics[width=0.9\linewidth]{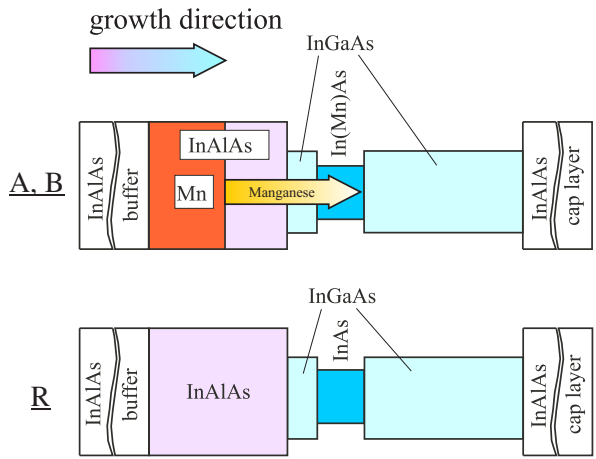}
\caption{Band diagrams of the QW structures under study excluding the
substrate and the InAlAs graded buffer.} \label{fig00}
\end{figure}

All experimental samples share the same QW design.
Following the band lineups  sketched on Fig.~\ref{fig00} an  In$_{0.75}$Ga$_{0.25}$As
shallow QW embedded in between In$_{0.75}$Al$_{0.25}$As barriers features an
asymmetrically inserted and compressively strained InAs channel of 4 nm.
 The distances between the InAs channel edges and the left and right InAlAs/InGaAs 
interfaces are 2.5 nm and 13.5 nm, respectively. Manganese containing samples
possess 7 nm layers of homogeneously Mn doped In$_{0.75}$Al$_{0.25}$As that are
inserted on the substrate side of the QW region and separated from the QW by an
In$_{0.75}$Al$_{0.25}$As barrier of 5 nm thickness.  Due to a segregation effect
during the MBE growth process a significant amount of manganese resides in the vicinity of
the InAs QW. Two structures, A and B, with different Mn concentration were realized.
The sample, referred as A in this paper, has a dopant concentration not exceeding 
 $n_{Mn} = 2\times 10^{20}$~cm$^{-3}$ that results in a Mn concentration of about two
orders of magnitude lower in the InAs QW as revealed by secondary ion mass spectrometry
(SIMS)~\cite{Wurstbauer2009}. Having an acceptor nature in III-V compounds Mn provides
a  two-dimensional hole gas (2DHG) in the InAs:Mn QW. Hall effect 
measurements determined a hole density as high as $10^{12}$~cm$^{-2}$ at $T = 4.2$~K.
As for the second structure B of the same design but of significantly lower doping concentration
(653$^\circ$C vs 852$^\circ$C  temperature of Mn effusion cell) conclusively a  two times
lower hole density was measured in magneto-transport experiments. In addition to these 
diluted magnetic semiconductors (DMS) QWs an intentionally undoped structure R without 
any Mn implementation was grown to serve as a reference.

\begin{figure}[b]
\includegraphics[width=0.7\linewidth]{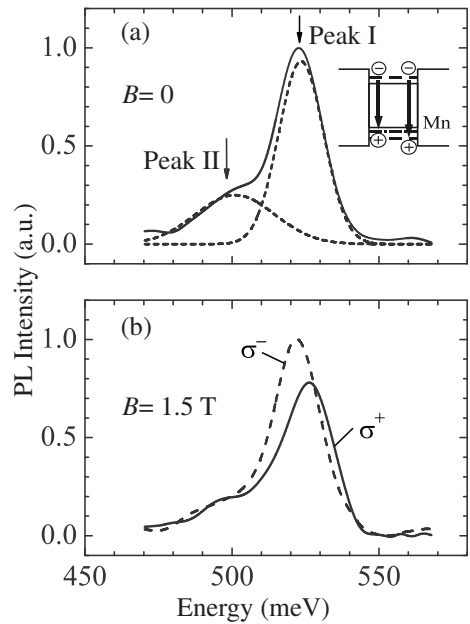}
\caption{PL spectra of sample A measured at $T$ = 2 K and
$W_{exc}$ = 10 W/cm$^2$. (a) No magnetic field is applied, (b)
circularly polarized magneto-PL spectra in the magnetic field of 1.5 T. The inset
shows schematically two paths of radiative transitions of photoelectrons in the Mn doped
QW.} \label{fig01}
\end{figure}

 \section{III. Experimental technique}

In our experiments we have used an experimental  setup designed to
measure polarization resolved magneto-PL in the infrared (IR) wavelength
regime ($2\div5$ $\mu$m).  The setup includes a magnetooptical
helium bath cryostat, a laser diode excitation source, an optical
polarization system and a grating or Fourier spectrometer equipped
with a nitrogen cooled IR photodetector.  Spectra are recorded
by using lock-in technique. Magnetic fields up to 6 T were applied
normally to the sample plain and along the wave vector of the
emission, i.e., the experiment is done in Faraday geometry.
The sample temperature can be varied from 2 up to 300 K. The
laser diode, operating in $cw$ mode, emits at wavelength $\lambda =
809$~nm and is focused onto a 1 mm diameter spot at the surface of the
sample. The excitation density $W_{exc}$ can be changed from 0.5 to
20~W/cm$^2$. The PL emission passes through a polarization system
consisting of a quarter wavelength retardation ZnSe Fresnel rhomb
and a linear polarizer having the optical axes crossed at an angle of $\pm$
45$^\circ$  to select $\sigma^+$ or $\sigma^-$ polarized light
~\cite{edge,Glazov2014}.

 \section{IV. Experimental results}

 All samples studied showed bright PL although the
signal from DMS samples was much weaker than that from
the reference structure.
The signal depends linearly on the excitation density in the explored range $W_{exc} = 0.5\div 20$~W/cm$^2$.

\begin{figure}[b]
\includegraphics[width=0.7\linewidth]{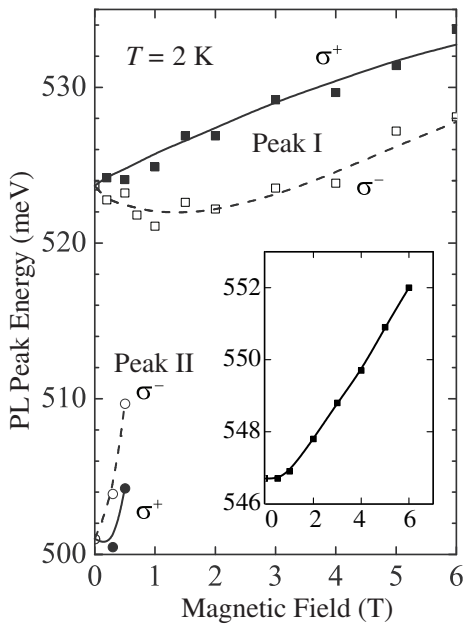}
\caption{Energy positions of $\sigma^+$ and $\sigma^-$ components of Peaks I and II as a function of the magnetic field 
measured in sample A at $T$ = 2 K for $W_{exc}$ = 10 W/cm$^2$.
The inset displays the corresponding dependence for structure R.
Lines are a guide for the eyes.} \label{fig3}
\end{figure}

\begin{figure}[t]
\includegraphics[width=0.7\linewidth]{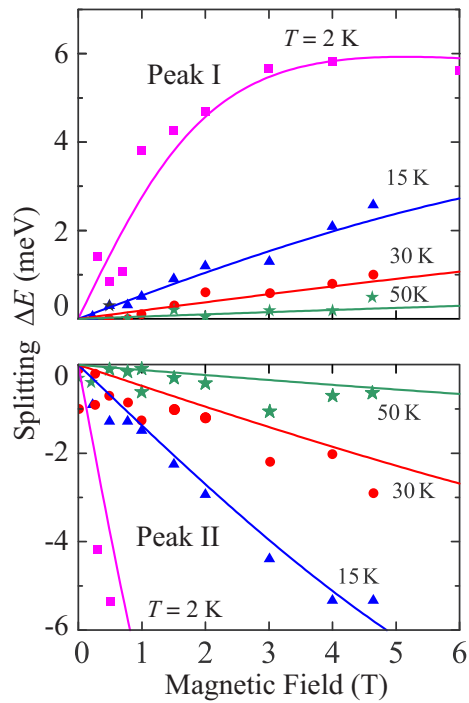}
\caption{Magnetic field dependence of the energy splitting
$\varDelta E$ of Peaks I and II, in sample A at different
temperatures. $\varDelta E$ is defined as $E_+-E_-$, where
$E_{+/-}$ is the energy corresponding to the $\sigma ^{+/-}$ polarized
components. Dots represent the experimental data obtained at different
temperatures. Lines correspond to
the theoretical fit after Eqs.~\eqref{eq:delta1} and
\eqref{eq:delta2}. The effective temperature $T_{Mn}$ is assumed to be equal to sample temperature $T$, except 
at $T=2K$, where the best fit corresponds to $T_{Mn}=3.3K$ } \label{fig4}
\end{figure}

The PL band of sample A exhibits  two contributions marked in
Fig.~\ref{fig01}(a) as Peak I and Peak II, which are separated
from each other by about 25 meV. In a magnetic field, both Peak I and
Peak II are blue shifted and split into two circularly polarized
components, Fig.~\ref{fig01}(b). Splitting reaches a large value up to 6 meV, Fig.~\ref{fig3}. At low
temperatures,  the magnetic field dependence of the energy splitting of Peak I
tends to saturate, see Fig.~\ref{fig4}. Increasing the temperatures above
15~K, we observe that the splitting of both peaks rapidly vanishes, Fig.~\ref{fig4}.

Peak I is $\sigma^-$ polarized whereas the polarization of Peak II is opposite. 
The degree of circular polarization of Peak I,  $P_{circ} = \frac{I_++ I_-}{I_+-
I_-}\times100\%$, where $I_{+/-}$~-- is the intensity of $\sigma^{+/-}$
polarized emission, reaches 30\% at $B = 6$ T.

Figure 5 shows the magnetic field dependence of the polarization of
both PL peaks at different temperatures. At low temperature, the polarization of Peak I linearly increases
with the magnetic field at small fields and saturates at high fields. With the temperature increase, the polarization decreases
and the saturation vanishes. Behavior of Peak II is similar though the polarization
at low temperatures can not be reliably determined due to the disappearance of Peak II in the magnetic fields stronger
than 1T.

Sample B with far low Mn doping also demonstrates two PL lines, Fig.~\ref{fig6}(a). Peak II is poorly
pronounced so that it cannot be analyzed. Peak I demonstrates a blue shift and a strong polarization
in the magnetic field ($P_{circ} = 40$\% at $B = 6$ T). The sign of the polarization is the same
as for sample A, but, in contrast to what was observed in sample A, Peak I in sample B exhibits no splitting
(Fig.~\ref{fig6}(b)) within our spectral resolution of  $\simeq$ 0.3 meV.

In reference sample R, PL contour consists of a single narrow peak (Peak I) while Peak II is absent,
Fig.~\ref{fig7}(a). Application of a magnetic field results in a blue shift of the PL line and a strong circular
polarization which reaches 54\% in a magnetic field of 6 T. Similar to sample B, no energy splitting of the peak has been detected
(Fig.~\ref{fig7}(b)).

\begin{figure}[t]
\includegraphics[width=0.7\linewidth]{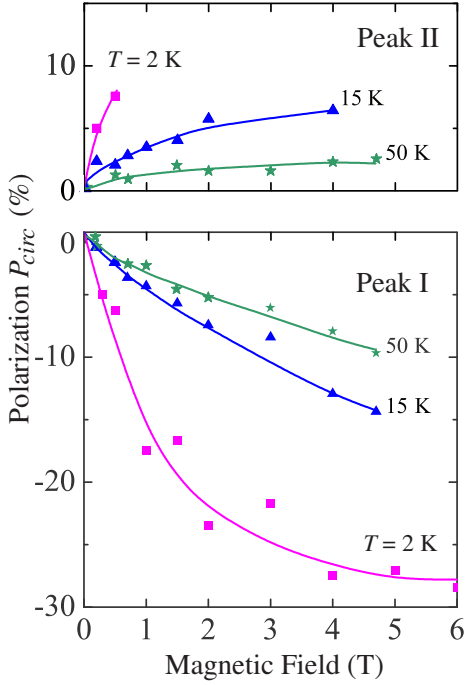}
\caption{Polarization degree $P_{circ}$ of both Peaks I and II
measured in sample A at $T$ = 2 K for $W_{exc}$ = 10 W/cm$^2$.
Lines are a guide for the eyes.} \label{fig5}
\end{figure}

 \begin{figure}[t]
\includegraphics[width=0.7\linewidth]{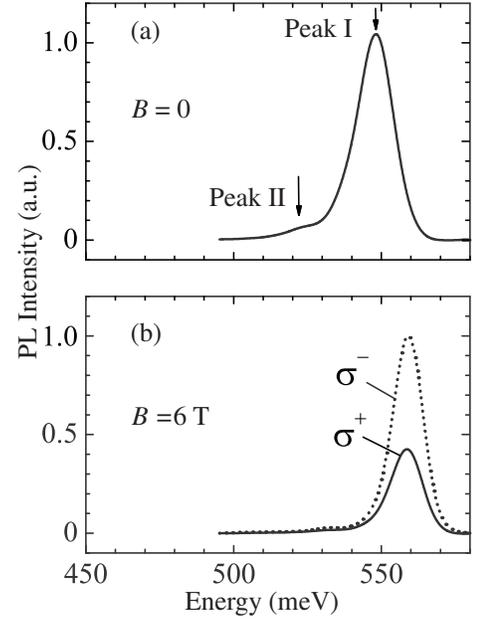}
\caption{PL spectra of sample B measured at $T$ = 2 K for
$W_{exc}$ = 10 W/cm$^2$. (a) without magnetic field, (b)
circularly polarized magneto-PL spectra obtained in a magnetic field of 6 T. }
\label{fig6}
\end{figure}

\begin{figure}[t]
\includegraphics[width=0.7\linewidth]{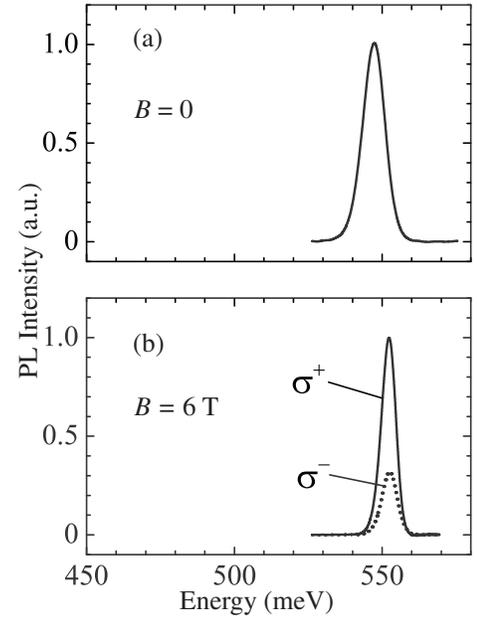}
\caption{PL spectra of sample R measured at $T$ = 2 K for
$W_{exc}$ = 10 W/cm$^2$. (a) without magnetic field, (b)
circularly polarized magneto-PL spectra for $B$ = 6 T. }
\label{fig7}
\end{figure}

The large energy spitting and polarization of the PL peaks measured in sample A
as well as their field and temperature dependencies point to the
important role of the exchange interaction between carriers confined in the QW
and Mn ions. An independent evidence for substantial exchange interaction 
has been obtained by studying of magneto-gyrotropic effect~\cite{3aa} in the
same structures~\cite{Olbrich2012}.

\section{V. Discussion}

We ascribe Peak I which dominates the PL signal of samples A, B as well
as the single PL peak observed from reference structure R to
radiative recombination of electrons and heavy holes
occupying the ground levels e1 and hh1 in the QW. It should be noted
that excitonic effects are negligible in our experiments. Indeed, the exciton binding
energy is known to be of the order of  1 meV in InAs QWs whereas
the Fermi energy due to the large hole concentration in studied Mn-doped samples is much larger.
The density of photoelectrons contributing to radiative recombination is far lower and they seem to be 
localized with binding energies of a few meV. Localization
centers can emerge due to inhomogenity of the InAs QW and presence of
charged ions. This conclusion is confirmed by the
quadratic magnetic field dependence of the PL peak energy, observed in
a magnetic fields up to $2\div3$ T for all studied structures, see Fig.~\ref{fig3} for sample A and the inset of 
Fig.~\ref{fig3} for sample R. In the case of sample A, the picture is complicated by the superimposed
non-linear on $B$ splitting, Fig.~\ref{fig3}. 


We attribute the red-shifted Peak II, detected in DMS structures, to
optical transitions of conduction electrons to the acceptor level of Mn ions,
embedded in the QW, see inset of Fig.~\ref{fig01}(a). Indeed,
a shallow manganese impurity band lying 23 meV below the InAs valence
band edge was revealed in bulk InMnAs~\cite{Chiu2004}. Using this
value as a reference for the Mn shallow acceptor binding energy in bulk InAs, we have calculated the binding energy $E_{Mn}$ in
the QW. For that we followed the method suggested in
Ref.~\cite{Averkiev} and treated the acceptor in zero-range potential approximation. 
Within the approach the bound state wavefunction $\Psi$ satisfies 
the Schr\"odinger equation:
\begin{equation}
\label{eqRozh1}
{H_L}\Psi = E_{Mn}\Psi+2\pi\delta \left( {{\bf{r}} - {{\bf{r}}_{\bf{0}}}} \right){\Psi _0},
\end{equation}
where $H_L$ is the Luttinger Hamiltonian, $\bf{r_0}$ is the acceptor position.
The 4-component function $\Psi _0$ is to be found from the boundary conditions.
At the impurity site $\mathbf{r_0}$ the boundary condition for the angular averaged wavefunction $\overline{\Psi}$ reads:
\begin{equation}
\label{eqRozh2}
\left.  \overline{\Psi}\right|_{\bf{r}\rightarrow\bf{r_0}}=\left(\frac{1}{|\bf{r}-\bf{r_0}|}-\alpha\right)\Psi_0+o(|\bf{r}-\bf{r_0}|),
\end{equation}
where $\alpha$ is the parameter of the impurity attractive potential strength in the zero-range potential model.
At the QW boundaries all components of $\Psi$ were set to zero thus implying the boundary conditions for 
the infinitely deep QW. Setting the QW width to infinity enabled us to find $\alpha$ from (\ref{eqRozh1}),(\ref{eqRozh2}) with $E_{Mn}$ being set to the Mn binding energy for the bulk. Restoring the QW width to its experimental value we obtained the binding energy $E_{Mn}$ from Eq.~(\ref{eqRozh1}) and~(\ref{eqRozh2}) with known $\alpha$. The
calculation shows that for the system studied, containing the 4 nm
wide QW, $E_{Mn}$ is practically the same as that in bulk
InAs, though in narrower QWs it decreases swiftly. The particular
value  $E_{Mn}\simeq$ 25 meV obtained for Mn ion in the center of the 4 nm QW is
in good agreement with the experimentally observed red shift of Peak II
relative to Peak I. Attribution of Peak II to the presence of Mn in QW is consistent
with the fact that Peak II is extremely weak in sample B having lower Mn content. 

We now turn to the case of an applied magnetic field. The magneto-PL of
the manganese-free  structure R was studied in detail in
Ref.~\cite{APL2014}. Here we give only a brief summary of the main
results which are important for the further discussion. The
application of a magnetic field results in the spin splitting of both,
conduction and valence bands. Optical recombination processes of
the electron $|e1,+1/2 \rangle$ with the hole $|hh1,-3/2 \rangle$
or the electron $|e1,-1/2 \rangle$ and the hole $|hh1,+3/2
\rangle$ are accompanied by the emission of $\sigma^-$- and
$\sigma^+$-polarized photons, respectively, that are detected in
the Faraday geometry of the experiment. The strong $\sigma^+$ circular polarization of
the magneto-PL peak stems from the spin polarization of the
nonequilibrium holes, characterized by a faster relaxation to the
ground Zeeman level in comparison with electrons. The same
polarization is observed in structures with degenerate 2DEG,
where the different electron spin states are equally
populated~\cite{APL2014}. The absence of the splitting of the PL line into
polarized components of different helicity was explained by the nearly equal
magnitude of the electron and hole Lande factors in the
system~\cite{APL2014}. Within the suggested model, the
polarization must invert its sign to $\sigma^-$ if we change
type of conductivity of the QW from $n$-type to strong $p$-type. Indeed, at the condition of equally
populated spin levels in the valence band in $p$-type samples, polarization is determined by electrons and 
recombination of electron at the ground state $|e1,+1/2 \rangle$ (for negative electron $g$-factor) with the hole $|hh1,-3/2 \rangle$ 
must dominate. This process is accompanied  by the emission of $\sigma^-$-polarized
photons. Samples A and B
contain degenerated 2DHG, thus $\sigma^-$ polarization  of 
Peak I detected from these samples is consistent with the model.
Zero splitting of circular polarized PL components in sample B should
be obviously attributed to vanishingly small Mn content in the
InAs QW.

In contrast to sample B, sample A provides clear evidence that free carriers interact via exchange 
with Mn ions embedded in the InAs QW. Indeed, application of a magnetic field results in
a large splitting of Peak I into circularly polarized components
(Fig.~\ref{fig4}) which is accurately described by the Brillouin
function in a wide temperature range (fitting details are given in
the next Section). The splitting tends to saturate in magnetic fields
higher than $3\div4$ T at the lowest achieved temperature of 2 K
and decreases rapidly if temperature is raised above $\sim$10 K.
The polarization dependence on magnetic field and temperature is similar to that of the splitting,
Fig.~\ref{fig5}.

To clarify the role of carrier heating by photoexcitation
we have measured the dependence of the polarization degree on
magnetic fields up to 6 T at different excitation power in the
range $5\div10$ W/cm$^2$ for sample A and $2.5\div10$ W/cm$^2$ for
sample B. In the first case no difference has been observed for different light intensities. 
For sample B the peak value of $P_{circ}$
at the lowest excitation level was 10\% higher than that for the maximum
excitation power while the shape of the curves did not change.
Thus the contribution of light heating can be considered as
minor.

We note that all the investigated structures show a polarization degree
which is considerably below 100\% even at $T = 2$~K and $B = 6$~T
although the dependence  $P_{circ}$ vs $B$ tends to saturate. This fact
can be ascribed to the interplay between the short lifetime of
photogenerated carriers compared to the spin relaxation time. In
the particular case of $p$-type QWs, where the holes are unpolarized
and the PL polarization is determined solely by the electron spin
polarization, the PL polarization is given by $P_c=-2S_e^{(0)}
\tau_r/(\tau_r + \tau_s)$. Here $S_e^{(0)}$ is the average
electron spin in thermal equilibrium, and $\tau_r$ and $\tau_s$ are
the electron lifetime  and spin relaxation time, respectively.

 \begin{figure}[t]
\includegraphics[width=0.7\linewidth]{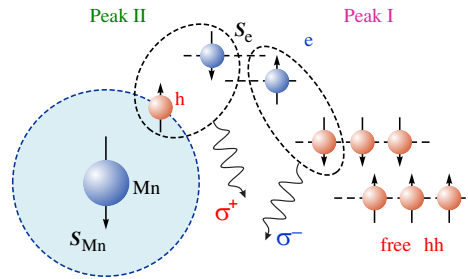}
\caption{\MD{Sketch of the optical transitions  contributing to
Peaks I and II in magnetic field. Only transitions contributing to dominant circularly polarized components of Peaks I and II
are depicted.}} \label{fig07}
\end{figure}


\subsection{Microscopic theory}

Now we discuss the microscopic model describing polarization and splitting of  Peaks I and II
in a magnetic field. Peak II is attributed to the optical
transitions between photoelectrons and holes bound to Mn ions (see
Fig.~\ref{fig07}) which can be schematically described as
\begin{equation}
\label{eq:peak1}
\mathrm e + (\mathrm h + \mathrm{Mn}) \rightarrow \gamma + \mathrm{Mn} \:,
\end{equation}
where the symbols $\rm e$, $\rm h$, and $\gamma$ denote  an
electron, hole, and photon, respectively.  Mn ion and the hole bound to Mn are antiferromagnetically coupled.
In the magnetic field, the bound-hole-Mn complex gets polarized in such a way that the hole spin points along the field direction.
Recombination of spin polarized holes with electrons leads to $\sigma^+$ circular polarization of Peak II.

Due to the strong $p$-$d$ exchange coupling, the ground state
of the bound-hole-Mn complex is described by the total angular
momentum $J = 1$ and is three-fold degenerate in the projection
$J_z$ at zero magnetic field~\cite{Averkiev1988}. A conduction-band electron, in contrast,
has spin $s_{e} = 1/2$, and each state is two-fold
degenerate. Thus, there are six different initial states for the
processes described by Eq.~\eqref{eq:peak1}. The final electron
state of Mn is six-fold degenerate in the projections of the Mn spin
$S_{\mathrm{Mn}} = 5/2$. The allowed optical transitions 
emitting $\sigma^+$-polarized radiation (labeled by the
index $k=1 \ldots 6$) and their relative intensities $C_k$ are
summarized in Table~\ref{tab1}. The transitions allowing the emission
of $\sigma^-$-polarized photons can be obtained from
Table~\ref{tab1} by simultanious inverting the sign of $J_z$,
$s_{e,z}$ and $S_{\mathrm{Mn},z}$.

\begin{table}
    \centering
        \begin{tabular}{c|c|c|c|l}
        \hline\hline
            & Initial state  & Final state & Relative & \multicolumn{1}{c}{Zeeman shift} \\
            $k$ &$J_z$, $s_{e,z}$ & $S_{{\rm Mn},z}$ & rate $C_k$& \multicolumn{1}{c}{$\Delta_k$} \\
            \hline
            1&-1, -1/2 & -5/2 & 1/2 & $-\Delta_e/2 - \Delta_1 + 5\Delta_{\mathrm{Mn}}/2$ \\
            2&-1, +1/2 & -3/2 & 1/10 & $\Delta_e/2 - \Delta_1 + 3\Delta_{\mathrm{Mn}}/2$ \\
            3&0, -1/2 & -3/2 & 1/5 & $-\Delta_e/2 + 3\Delta_{\mathrm{Mn}}/2$ \\
            4&0, +1/2 & -1/2 & 1/10 & $\Delta_e/2 + \Delta_{\mathrm{Mn}}/2$ \\
            5&+1, -1/2 & -1/2 & 1/20 & $-\Delta_e/2 + \Delta_1 + \Delta_{\mathrm{Mn}}/2$ \\
            6&+1, +1/2 & +1/2 & 1/20 & $\Delta_e/2 + \Delta_1 - \Delta_{\mathrm{Mn}}/2$ \\
            \hline \hline
        \end{tabular}
        \caption{Optical transitions between conduction-band electrons and holes
        bound to Mn with emission of $\sigma^+$-polarized photons.
        $\Delta_e$, $\Delta_1$, and $\Delta_{\Mn}$ are the Zeeman splitting of electron states, hole-Mn complex, and Mn ion, respectively.}
        \label{tab1}
\end{table}

In an external magnetic field, the emission line splits  into
12 components, each of them being either $\sigma^+$- or
$\sigma^-$-polarized. The corresponding Zeeman shifts of the
$\sigma^+$-polarized components are listed in Table~\ref{tab1}. In a
real QW structure, however, individual components may not be
spectrally resolved due to considerable inhomogeneous broadening.
Therefore, the measured spectrum of the $\sigma^+$-polarized PL is
given by
\begin{equation}\label{eq:spectrum}
I_+(\hbar \omega) = \sum \limits_{k=1}^6 C_k f_k D(\hbar \omega - \Delta_k)\:,
\end{equation}
where $f_k$ is the population of the initial state with the  index
$k$ and $D(\hbar \omega)$ is the PL contour at zero magnetic
field. To first order in the Zeeman splitting,
Eq.~\eqref{eq:spectrum} is equivalent to
\begin{equation}\label{eq:spectrum2}
I_+(\hbar \omega) \approx D(\hbar \omega) \sum \limits_k C_k f_k - D'(\hbar\omega) \sum \limits_k C_k \Delta_k f_0 \:,
\end{equation}
where $D'(\hbar\omega) = d D(\hbar\omega) / d \hbar \omega$ and
$f_0$  is the population at zero magnetic field, identical for all
initial states. Equation~\eqref{eq:spectrum2} describes that the PL contour
$I_+(\hbar \omega)$ is spectrally shifted with respect to the PL
contour at zero field by $\sum_k C_k \Delta_k / \sum_k C_k$.
Similarly, the $\sigma^-$-polarized PL contour is shifted in the
opposite direction by the same value. For the particular case of
the optical transitions listed in Table~\ref{tab1}, the effective
Zeeman shift between the broadened  PL lines of opposite helicity is
given by
\begin{equation}\label{eq:delta1}
\Delta E_{II} = -\Delta_1 - \frac12 {\Delta_e} + \frac52 \Delta_{\Mn} \:.
\end{equation}

The Zeeman splitting of the hole-Mn complex, Mn ion, and conduction-band electron have the form
\begin{equation}
\label{}
\Delta_1 = g_1 \mu_0 B_z\:,\;\; \Delta_{\Mn} = g_0 \mu_0 B_z\:,
\end{equation}
\[
\Delta_e = g_e \mu_0 B_z + a\,\mathrm{B}_1\left( \frac{g_1 \mu_0 B_z}{k_B T_{Mn}} \right),
\]
where $g_1$ is the $g$-factor of the hole-Mn complex, $g_0 = 2$ is
the Mn $g$-factor, $g_e$  is the intrinsic electron $g$-factor,
$\mu_0$ is the Bohr magneton, $T_{Mn}$ is the effective
temperature of the Mn spins, and $\mathrm B_1(x) = 2 \sinh x/(1 + 2\cosh
x)$ is the Brillouin function of the momentum $J = 1$. The second
contribution to $\Delta_e$ describes the splitting due to exchange
interaction between conduction-band electrons and hole-Mn
complexes, parameter $a$ depends on the exchange interaction
strength and complex concentration.

The PL Peak I is attributed to the optical transitions between
electrons and  heavy holes confined in the QW, see Fig.~\ref{fig07}.
The sign of the PL circular polarization in a magnetic field is
largely determined by the spin polarization of electrons since the
thermal spin polarization of holes is low in $p$-doped structures.
The electron spin polarization can occur due to thermal population
of spin-split states as well as spin-dependent extraction of
electrons caused by other recombination
channels~\cite{Korenev2012}, e.g., radiative transitions
contributing to Peak II. The spectral shift between the
polarized lines of opposite helicity in a magnetic field is determined by the
Zeeman splitting of both, conduction and valence bands, and is given
by
\begin{equation}
\label{eq:delta2}
\Delta E_{I} = -\Delta_e + \Delta_{hh}\:,
\end{equation}
where $\Delta_{hh}$ is the Zeeman splitting of the heavy-hole subband,
\begin{equation}
\Delta_{hh} = g_{hh} \mu_0 B_z + b\,\mathrm{B}_1\left( \frac{g_1 \mu_0 B_z}{k_B T_{Mn}} \right),
\end{equation}
Here, $b$ is the parameter describing the strength of the exchange
interaction  between free holes and hole-Mn complexes.

Peak I and its splitting in magnetic fields is well observed
in experiment. In the regime of small
magnetic fields, $\Delta E_{II}$ depends linearly on the magnetic
field and has two contributions: one is temperature independent and the other one scales as $1/T_{Mn}$.
 Fitting the experimental data presented in
Fig.~\ref{fig4} by Eq.~\eqref{eq:delta2} yields $(b-a)g_1 \approx
24$~meV and $g_e - g_{hh} \approx 2.6$. The small absolute value
of $g_e - g_{hh}$ is in agreement with negligible splitting of the
PL line in QW structures without magnetic
impurities~\cite{APL2014}. We note that at temperatures $T < 5$~K, PL
spectra are weakly sensitive to the sample temperature. This indicates that the effective Mn temperature
$T_{Mn}$ is higher then the nominal sample temperature due to heating by radiation. In particular, the
best agreement between the experimental curve measured at 2~K and
theory is obtained for an effective temperature $T_{Mn}
\approx 3.3$~K. Peak II is less pronounced and disappears in high
magnetic fields, therefore its treatment is less reliable. Fitting
the data on the Zeeman splitting of Peak II by
Eq.~\eqref{eq:delta1} yields $a g_1 \approx -126$~meV and $2g_1 +
g_e \approx -4.6$.

The saturation of the splitting at higher magnetic fields is
determined by $g_1$. The best fit of the experimental data for Peak
I using $T_{Mn} = 3.3$~K is obtained for $g_1 \approx 3.5$. This value is in agreement with 
the theoretical calculation
 for the $g$-factor of a bound-hole-Mn complex~\cite{Averkiev}. Taking this $g_1$ value, all
other parameters can be estimated as $a \approx -36$~meV, $b
\approx -29$~meV, $g_e \approx -11.6$, and $g_{hh} \approx -14.2$
for the QW structure under study. The negative sign of $b$ is in
agreement with the antiferromagnetic behavior of the $p$-$d$
exchange interaction. The determined negative sign of the parameter $a$,
which describes the exchange interaction between the
conduction-band electron and bound-hole-Mn complex, may be caused by
the electron-hole exchange interaction and is discussed in
Ref.~\onlinecite{Sliwa2008}.

Surprisingly, the observed splitting of the PL lines in sample A is two orders of magnitude larger
than estimated  if one
uses the $s-d$ and $p-d$ exchange-coupling constants for bulk InMnAs,
$\alpha$ and $\beta$  of  order of -1 eV and 0.5 eV, respectively, 
\cite{Zudov2002}. One possible explanation for this amazing fact is the inhomogeneous Mn
distribution in the plane of the QW. Indeed, even in the case of high quality metamorphic buffer layers used in our
 structures the density of extended defects (threading dislocations) lies around $10^{6}$~cm$^{-2}$,
 maximum $10^{7}$~cm$^{-2}$. 
Due to well-known phenomenon --- enhanced metal diffusion and accumulation along the threading dislocations in III-V 
semiconductors~\cite{Liliental2002,Ivanov2004} --- Mn atoms can accumulate around the residual threading dislocations 
propagating into the QW. Thus, their local concentration in the InAs Qw can exceed the average level of segregated Mn concentration
$\sim 10^{18}$~cm$^{-3}$ and may in principle be high enough to create the regions of InMnAs with Mn content on the order of 1 \%.

\section{VI. Summary}

To summarize, we have demonstrated that strained InAs QWs with
low concentration of manganese ($\sim$0.01$\%$) exhibit
bright low-temperature photoluminescence. It is shown that the Zeeman
effect in such strongly diluted magnetic heterostructures can be
accurately investigated by using magneto-PL as a highly informative 
method. An anomalously giant splitting of the PL lines in magnetic fields was observed in
one of the diluted magnetic InAs QW structures.
This surprising finding is ascribed to local Mn accumulation along threading dislocations.

Financial support by the  DFG (SFB~689), Russian Foundation 
for Basic  Research, RF President grant MD-3098.2014.2, 
and EU project SPANGL4Q is gratefully acknowledged.

{

\end{document}